\documentclass[pra,fleqn,showpacs]{revtex4}

\begin{document}

\title{Nonrelativistic ionization energy for the helium ground state.}
\author{Vladimir I.~Korobov}
\affiliation{Joint Institute for Nuclear Research\\
141980, Dubna, Russia}
\email{korobov@thsun1.jinr.ru}
\pacs{31.15.Pf, 31.15.Ar}

\begin{abstract}
The helium ground state nonrelativistic energy with 24 significant digits
is presented. The calculations are based on variational expansion with
randomly chosen exponents. This data can be used as a benchmark for other
approaches for many electron and/or three-body systems.
\end{abstract}

\maketitle

From the early days of the quantum mechanics the ground state ionization
energy of the helium atom was a benchmark for approximate methods of
solving nonrelativistic Schr\"odinger equation for a few--body system. One
of the earliest variational calculations has been performed by Hylleraas
\cite{Hylleraas} in 1929 and it yields an about 5 significant digits for
the nonrelativistic ionization potential. In 1957, Kinoshita
\cite{Kinoshita} presented a 7 digit number obtained with a 39 parameter
function, which along with higher order corrections including the Lamb
shift calculations confirmed a very good agreement with the best
experimental value. Since that time with the development of computer power
the accuracy grows very rapidly. We would like to mention here the two
most recent calculations. The first \cite{Sims} is aimed to elaborate an
efficient variational method for the many electron atoms. The second
\cite{Drake} is to find an effective and economical way for studying the
helium and helium-like two electron atoms.

In this short report we want to present a new very accurate value for the
nonrelativistic energy of the helium ground state. In our calculations we
strictly follows a method described in \cite{var00}. The two modifications
are to be stated. First, a sextuple precision arithmetics (about 48
decimal digits) implemented as a Fortran-90 module has been used instead
of a multiprecision package written by Bailey. The use of this module
gives an about 5-fold improvement in computational time and allows for to
increase significantly a length of a basis set. The module is based on a
representation of a sextuple precision number by a set of 3 double
precision numbers. It is assumed that an {\em exact} sum of these double
precision numbers is some sextuple precision number \cite{sextuple}.
Second, we have taken a multilayer variational wave function with 5
independent sets of variational parameters (instead of 4 as in
\cite{var00}), which consecutively approximates one after another smaller
and smaller distances of electrons with respect to a nucleus.

A variational wave function is expanded in a form \cite{var00}
\begin{equation}
\begin{array}{@{}l}
\displaystyle
\psi_0 = \sum_{i=1}^{N/2}
\Big\{U_i\,\mbox{\sl Re}
\bigl[\exp{(-\alpha_i r_1-\beta_i r_2-\gamma_i r_{12})}\bigr]
+W_i\,\mbox{\sl Im}
\bigl[\exp{(-\alpha_i r_1-\beta_i r_2-\gamma_i r_{12})}\bigr]
\Big\}
\end{array}
\end{equation}
where $\alpha_i$, $\beta_i$ and $\gamma_i$ are complex parameters
generated in a quasi-random manner:
\begin{equation}\label{random}
\alpha_i = \left[\left\lfloor\frac{1}{2}i(i+1)\sqrt{p_\alpha}
\right\rfloor(A_2-A_1)+A_1\right]
+i\left[\left\lfloor\frac{1}{2}i(i+1)\sqrt{q_\alpha}
\right\rfloor(A'_2-A'_1)+A'_1\right],
\end{equation}
$\lfloor x\rfloor$ designates the fractional part of $x$,
$p_\alpha$ and $q_\alpha$ are some prime numbers, $[A_1,A_2]$ and
$[A'_1,A'_2]$ are real variational intervals which need to be
optimized. Parameters $\beta_i$ and $\gamma_i$ are obtained in a
similar way. The actual values of these parameters for the calculation
with the largest basis set of 5200 functions is presented in Table I. As
is seen from the Table fine tuning of variational parameters is not
required that greatly facilitates calculations.

Table II demonstrates a convergence of the variational expansion with the
number of basis functions. Extrapolated value has been obtained by
means of the simple extrapolation formula,
\begin{equation}
E(\text{extrap}) = E(N)-C\times N^{-\nu},
\end{equation}
where parameters $C$ and $\nu$ are taken from the best fit of the last
4 or 5 calculations.

In Table III a comparison with the most recent and most accurate values is
presented. Our result extends accuracy of the previous calculations by
more than three orders of magnitude.

The author would like to thank J.S.~Sims and G.W.F.~Drake for stimulating
to publish this work.


\newpage

\begin{table}
\caption{Variational parameters and number of basis functions ($n_i$) for
different subsets of the variational wave function with $N=5200$.
Intervals $[A_1,A_2]$ and $[A'_1,A'_2]$ correspond to real and imaginary
parts of a randomly chosen parameter $\alpha_i$ (see Eq.~(\ref{random})
for details), intervals $[B_1,B_2]$ and $[B'_1,B'_2]$ to $\beta_i$, and
intervals $[G_1,G_2]$ and $[G'_1,G'_2]$ to $\gamma_i$. Prime numbers are
$p_\alpha=2$, $p_\beta=3$, $p_\gamma=5$, and $q_\alpha=7$, $q_\beta=11$,
$q_\gamma=13$.}
\begin{center}
\begin{tabular}{c@{~~~~}r@{~~~~~}r@{~~}r@{~~}r@{~~}r@{~~~~}
r@{~~}r@{~~}r@{~~}r@{~~~~}r@{~~}r@{~~}r@{~~}r}\hline\hline
 & $n_i~$ & $A_1$ & $A_2$ & $A'_1$ & $A'_2$ & $B_1$ & $B_2$ & $B'_1$ & $B'_2$
 & $G_1$ & $G_2$ & $G'_1$ & $G'_2$ \\
\hline
$i=1$ & 1160 & 0.70 & 3.10 & 0.00 & 0.60 & 0.95 & 3.35 & 0.00 & 0.50 & 0.00 & 0.80 & 0.00 & 0.75 \\
$i=2$ & 1160 & 1.00 & 9.35 & 0.00 & 1.05 & 0.60 & 7.60 & 0.00 & 0.70 & 0.00 & 0.70 & 0.00 & 1.80 \\
$i=3$ & 1020 & 3.00 & 14.0 & 0.00 & 0.00 & 3.50 & 14.0 & 0.00 & 0.00 & 0.00 & 1.60 & 0.00 & 1.60 \\
$i=4$ &  950 & 16.0 & 48.0 & 0.00 & 0.00 & 16.0 & 48.0 & 0.00 & 0.00 & 0.00 & 10.0 & 0.00 & 5.50 \\
$i=5$ &  910 & 45.0 & 200. & 0.00 & 0.00 & 45.0 & 200. & 0.00 & 0.00 & 0.00 & 30.0 & 0.00 & 11.0 \\
\hline\hline
\end{tabular}
\end{center}
\end{table}

\begin{table}[h]
\caption{Nonrelativistic energies for the ground state of a helium atom
$^\infty\mbox{He}$. $N$ is the number of basis functions. The last digits
of the difference in energy between two successive calculations is shown
in a third column.}
\begin{center}
\begin{tabular}{llr}
\hline\hline
$~~N~~~~~$ & $~~~~~~~~~E_{nr}$ (in a.u.) & $~\Delta E$ \\
\hline
3400   & $-$2.9037243770341195983110931 & \\ 
3800   & $-$2.9037243770341195983111421 &  490 \\
4200   & $-$2.9037243770341195983111540 &  119 \\
4600   & $-$2.9037243770341195983111572 &   32 \\
5200   & $-$2.9037243770341195983111587 &   15 \\
\hline
extrap & $-$2.9037243770341195983111594(4) \\
\hline\hline
\end{tabular}
\end{center}
\end{table}

\begin{table}[h]
\caption{Comparison with other calculations.}
\begin{center}
\begin{tabular}{lr@{\hspace*{3mm}}l}
\hline\hline
                            & $N$~& \hfil$E$ (a.u.) \\ \hline
Goldman \cite{Goldman}      & 8066  & $-$2.903724377034119594    \\
Chuluunbaatar  \\[-1mm]
~~~~{\it et al.} \cite{Chuka} & 5669  & $-$2.90372437703411959829  \\
Sims and \\[-1mm]
~~~~Hagstrom\cite{Sims}       & 4648  & $-$2.9037243770341195982999 \\
Drake {\it et al.} \cite{Drake}
                            & 2358  & $-$2.903724377034119598305  \\
                            & extrap& $-$2.903724377034119598311(1) \\[1mm]
This work                   & 5200  & $-$2.903724377034119598311159 \\
\hline\hline
\end{tabular}
\end{center}
\end{table}


\begin{thebibliography}{99}
\bibitem{sextuple} this version of the sextuple precision module is
  written by the author of this work and can be obtained upon request.
\end{thebibliography}
\end{document}